\documentclass[prd,twocolumn,showpacs,preprintnumbers,amsmath,amssymb]{revtex4}
\usepackage{graphicx}
\usepackage{dcolumn}
\usepackage{bm}
\usepackage[T1]{fontenc}
\usepackage{amssymb, latexsym, amsopn}

\newcommand{\av}[1]{\langle{#1}\rangle}
\newcommand{\avi}[1]{\langle{#1}\rangle_i}

\begin{document}

\title{Bayesian analysis of the backreaction models}

\author{Aleksandra Kurek}
\affiliation{Astronomical Observatory, Jagiellonian University, Orla 171, 30-244 Krak{\'o}w, Poland}
\email{alex@oa.uj.edu.pl}

\author{Krzysztof Bolejko}
\affiliation{Department of Mathematics and Applied Mathematics, \\
University of Cape Town, Rondebosch 7701, South Africa}
\affiliation{Nicolaus Copernicus Astronomical Center, Bartycka 18, 00-716 Warsaw, Poland}
\email{bolejko@camk.edu.pl}

\author{Marek Szyd{\l}owski}
\affiliation{Mark Kac Complex Systems Research Centre, Jagiellonian University, Reymonta 4, 30-059 Krak{\'o}w, Poland}
\affiliation{Dipartimento di Fisica Nucleare e Teorica, Universit{\`a} degli studi di Pavia, via A. Bassi 6, I-27100 Pavia, Italy}
\email{uoszydlo@cyf-kr.edu.pl}

\date{\today}

\begin{abstract}
 We present the Bayesian analysis of four different 
types of backreation models, which are based on the Buchert equations.
In this approach, one considers a solution to 
the Einstein equations for a general matter distribution and 
then an average of various observable quantities is taken.
Such an approach became of considerable interest when it
was shown that it could lead to agreement with
observations without resorting to dark energy. 
 In this paper we compare the ${\Lambda}$CDM model and the backreation models
 with SNIa, BAO, and CMB data, and find that the former is favoured.
However, the tested models were based on some particular assumptions 
about the relation between the 
average spatial curvature and the backreaction,
as well as the relation between the curvature and curvature index.
In this paper we modified the latter assumption, leaving the former unchanged.
We find that, by varying the relation between the curvature and curvature index,
we can obtain a better fit. 
Therefore, some further work is still needed -- in particular
the relation between the 
backreaction and the curvature
should be revisited in order to fully determine
the feasibility of the backreaction models to mimic dark energy.
\end{abstract}


\maketitle

\section{Introduction}
The Universe, as observed, is very inhomogenous on almost all scales.
However, in a standard approach to cosmology, it is assumed that the Universe can be described by the homogeneous and isotropic 
Friedmann--Lema\^itre--Robertson--Walker (FLRW) models. 
The FLRW models provide a remarkably precise description of cosmological observations, but to achieve this we need to pay a price --
in order to obtain concordance with observations 
it must be assumed that the Universe is filled with an unknown substance called dark energy. However, this substance has never been observed directly
and, since it has very unusual properties, some have begun to ask whether dark energy is real or if it is the description of the Universe which, requires the existence of such an exotic entity, that is invalid.

While it is possible that our Universe is
filled with dark energy, many alternatives have been proposed:
brane-world cosmologies (see \cite{branes-review} for a review),
f(R) cosmology (see \cite{f(R)-review} for a review), application of inhomogeneous
cosmological models (for a review see \cite{icm-review}) and others.
One of the recently proposed approaches is 
 based on an averaging framework.
Such an approach is motivated by the fact that the Einstein 
equations are non-linear, which means that 
the solution of the Einstein equations for a homogeneous matter distribution is
different from the averaged solution to 
the Einstein equations for a general matter distribution.
In other words,
the evolution of the homogeneous model might be slightly  different
from the evolution of an inhomogeneous Universe, even though inhomogeneities in the Universe, when averaged  
over a sufficiently large scale, might tend to be zero.
The difference between the evolutions of 
homogeneous and inhomogeneous models of the Universe is known
as the backreaction effect. 
In this approach, one considers a solution to 
the Einstein equations for a general matter distribution and 
then an average of various observable quantities is taken. 
Under certain assumptions such an attempt leads
to the Buchert equations \cite{B00}. The Buchert equations are very similar to the Friedmann equations, except for the backreaction term, which is
in general nonvanishing if inhomogeneities 
are present.
For a review on the backreaction effect and the
Buchert averaging schemem the reader is referred to
\cite{R06,B08,BC08}.  
Based on this scheme, Larena et al. have 
recently proposed a model \cite{LABKC08} in which
the metric of the Universe at a given instant looks
like the FLRW metric, but the  evolution of the scale factor is governed by the Buchert equations.
In this paper we aim to perform the Bayesian analysis of the cosmological observations within the models proposed 
in \cite{LABKC08}.

\section{Backreaction models}\label{model}

If the averaging procedure is applied to the Einstein equations then for
irrotational, pressureless matter and
3+1 ADM space-time foliation with a constant lapse and a 
vanishing shift vector, the following equations are obtained
\cite{B00}

\begin{eqnarray}
&& 3 \frac{\ddot{a}}{a} = - 4 \pi G \av{\rho} + \mathcal{Q},
\label{bucherteq1} \\ &&  3 \frac{\dot{a}^2}{a^2} = 8 \pi G \av{ \rho}	-
\frac{1}{2} \av{ \mathcal{R} } - \frac{1}{2} \mathcal{Q}, \label{bucherteq2} \\
&&  \mathcal{Q} \equiv \frac{2}{3}\left( \av{{\Theta^2}} - \av{ \Theta }^2 \right)
- 2 \av{ \sigma^2},
\label{qdef}
\end{eqnarray}
where a dot ($\dot{}$) denotes $\partial_t$,
$\av{ \mathcal{R} }$ is an average of the spacial Ricci scalar $^{(3)}
\mathcal{R}$, $\Theta$ is the scalar of expansion, $\sigma$ is the shear
scalar, $\rho$ is the matter density, and $\av{\ }$ is the volume average over the hypersurface of constant time:
$\av{A} = ( \int d^3x \sqrt{-h} )^{-1} \int d^3x \sqrt{-h} A $.
The scale factor $a$ is defined as a cube root of the volume:

\begin{equation}
a = \left( \frac{V}{V_0} \right)^{1/3},
\label{aave}
\end{equation}
where V$_0$ is an initial volume.

Equation (\ref{bucherteq1}) is compatible with  (\ref{bucherteq2}) if the
following integrability condition holds

\begin{equation}
\frac{1}{a^6} \partial_t \left( \mathcal{Q} a^6 \right) + 
\frac{1}{a^2} \partial_t \left( \av{R} a^2 \right)  = 0.
\label{intcond} 
\end{equation}
Similarly, as in the FLRW models, the following parameters can be introduced:

\begin{equation}
H \equiv \frac{\dot{a}}{a}, \quad \Omega_m \equiv \frac{8 \pi G}{3 H^2} \av{\rho}, \quad
\Omega_{\mathcal{R}} \equiv - \frac{\av{\mathcal{R}}}{6H^2},
\quad \Omega_{\mathcal{Q}} \equiv  - \frac{\av{\mathcal{Q}}}{6H^2}.
\end{equation}
The Hamiltonian constraints can then be written as:

\begin{equation}
\Omega_m  + \Omega_{\mathcal{R}} + \Omega_{\mathcal{Q}} = 1. 
\end{equation}
Observe that $\Omega_{\mathcal{R}} + \Omega_{\mathcal{Q}}$
can act like $\Omega_{\Lambda}$.
Moreover, 
if the dispersion of the expansion is
large then $\mathcal{Q}$ can be large and as seen from  (\ref{qdef}),  
one can get acceleration ($\ddot{a}>0$) without the need for dark energy.

The template metric of the Universe - the metric which describes 
the averaged universe can be written as

\begin{equation}
{\rm d} s^2 = {\rm d} t^2 - \frac{a(t)^2}{1 - k(t) r^2} {\rm d} r^2 - a(t)^2 r^2 \left( {\rm d} \vartheta^2 
+ \sin^2 \vartheta {\rm d} \varphi^2 \right).
\label{thm}
\end{equation}
A similar approach, i.e. to consider the template metric with a scale factor which evolves accordingly to the Buchert equations instead of 
the Friedmann equations, was first introduced by 
Paranjape and Singh \cite{PS08}, though in their model $k$ was constant.
The motivation for $k(t)$ comes from the fact that 
the averaged spatial curvature, if calculated 
at one instant, does not have to be the same as 
the averaged spatial curvature calculated at another instant. 
This is closely related to the fitting problem
closely studied by Ellis and Stoeger \cite{ES87}.
In considering the fitting problem, it becomes apparent that 
a homogeneous model fitted to inhomogeneous data can evolve quite differently 
from the real Universe.
Therefore, if inhomogeneous model is averaged at one instant
its FLRW parameters may be different for the FLRW parameters
obtained after averaging at another instant.

The Buchert equations do not form a closed system.
To close these equations and thus to calculate the evolution of the scale
factor one has to introduce some further assumptions \cite{B00}. 
One such assumption can be:
$\av{{\cal R}} \sim {\cal Q}$ \cite{LABKC08}.
As seen from the integrability condition (\ref{intcond}), this leads to

\begin{equation}
\av{\mathcal{R}} = \avi{\mathcal{R}} a^n \quad {\rm ~and~} \quad
\mathcal{Q} = -\frac{n+2}{n+6} \avi{\mathcal{R}} a^n,
\label{ansatz1}
\end{equation}
where $n$ is an arbitrary parameter.
Now, the final step is to derive a relation between the average
spacial curvature $\av{\mathcal{R}}$ and the
curvature index $k$. In analogy
to the FLRW models, the following relation can be proposed \cite{LABKC08}:

\begin{eqnarray}
&& k =   \frac{a^2 \av{\mathcal{R}}}{a^2_i |\avi{\mathcal{R}}|}, 
\nonumber \\
&& \rightarrow  \quad k(z) = 
- \frac{(n+6)(1 - \Omega_m) (1+z)^{-(n+2)}}{ |(n+6)(1-\Omega_m) |}.
\label{ansatz2}
\end{eqnarray}
In Sec.~\ref{curidx} we will modify the above assumption
and test models with different relations between $k$ and $\av{\mathcal{R}}$.
Summarizing, the model considered in this paper
is described by the metric (\ref{thm}), but the evolution 
of the scale factor is governed by the Buchert equations.
Employing the assumptions (\ref{ansatz1}) and (\ref{ansatz2})
the evolution equations reduce to the following relation:

\begin{equation}\label{hubble}
H = H_0 \sqrt{\Omega_m (1+z)^3 + (1-\Omega_m) (1+z)^{-n}}.
\end{equation}
This model is parametrized by two parameters:
$\Omega_m$ and $n$.
The distance, using (\ref{thm}), can then be calculated by solving 

\begin{equation}
\frac{d r}{dz} = \sqrt{ \frac{1 - kr^2}{\Omega_m (1+z)^3 + (1-\Omega_m) (1+z)^{-n}}}.
\label{dis}
\end{equation}

Larena et al. \cite{LABKC08} tested this model with a likelihood analysis
using the supernova and CMB data. They found that this
model is in good agreement with observations.
In the next section we will perform the Bayesian analysis of this model
using the type Ia supernovae (SNIa) data, baryon acoustic oscillations (BAO) and the 
observation of the cosmic microwave background (CMB) radiation.

\section{Bayesian analysis}\label{bayan}

The model presented in the preceding section 
will be confronted with cosmological observations in the Bayesian framework
using the CosmoNest code \cite{Mukherjee:2005wg}
\footnote{The CosmoNest code uses the nested sampling algorithm \cite{Skilling} and is a part of the CosmoMC code \cite{Lewis:2002ah}.}
which was adapted to our case.
In the Bayes theory all that we know about the vector of parameters ($\bar{\theta}$) of a given model ($M$) 
is contained in the posterior probability density function (PDF), which is given by \cite{J61}

\begin{equation}\label{posterior}
P(\bar{\theta}|D,M)=\frac{P(D|\bar{\theta},M)P(\bar{\theta}|M)}{ P(D|M)  },
\end{equation}

where  $D$ denotes the set of data used in the analysis;
$P(D|\bar{\theta},M)$ is the likelihood function for a given model,
which will henceforth be referred to as $L(\bar{\theta})$;
 $P(\bar{\theta}|M)$ is the prior PDF, which enables 
us to include our previous knowledge (i.e. without information coming from the data $D$) about the parameters under consideration;
 and the last quantity $P(D|M)$  is the normalization constant,
called the evidence (or marginal likelihood), which 
is the most important quantity in the Bayesian framework of model comparison.
The posterior PDF could be simply summarized in terms of a best fit value, which could be the posterior PDF mode (the most probable value of $\bar{\theta}$) or the mean of the marginal posterior PDF of a given parameter ($\theta_i$), obtained by integration (\ref{posterior}) over the remaining parameters.

\subsection{Parameters estimation}\label{parest}

\subsubsection{Supernova data}

\noindent
Firstly, we consider observations of Type-Ia supernova, which are taken from the 
 Supernova Legacy Survey \cite{A06} (SNLS) and the Union Supernova Compilation \cite{K08}.
After analytical marginalization over the $H_0$ parameter the likelihood function 
is of the following form

\begin{equation}
  L_{SN}(\Omega_{m},n) \propto \exp \left[-\frac{1}{2}\left(\sum_{i=1}^{N}\frac{x_i^2}{\sigma_i^2}-\frac{\left(\sum_{i=1}^{N}\frac{x_i}{\sigma_i^2}\right)^2}{\sum_{i=1}^{N}\frac{1}{\sigma_i^2}}\right)\right],
\end{equation}
where $N$ is the number of data points ($N=115$ for the SNLS sample and $N=307$ for the Union sample), $\sigma_{i}$ is the observational error \footnote{SNLS error estimation includes:  photometric uncertainty, uncertainty due to host galaxy peculiar velocities of 300 km/s, uncertainty of 0.13104 mag
related to intrinsic dispersion of SNe Ia. 
UNION statistical uncertainty was obtained from the light-curve fit.}, $x_i=\mu_{i}^{\mathrm{theor}}-\mu_{i}^{\mathrm{obs}}$, with $\mu_{i}^{\mathrm{obs}}=m_{i}-M$
($m_{i}$ is the apparent magnitude and $M$ the absolute magnitude of SNIa), 
is the observed distance 
moduli \footnote{The distance estimator was assumed to be 
   $\mu^{obs}=m_{B}- M+\alpha(s-1)-\beta c$. Where $m_{B}$ (supernova B-band
maximum magnitude), $s$  (stretch), and $c$ (color) were derived from the fit
to the light curves.
We took the observational data at face value without correcting
it for such effects as gravitational lensing.
}, 
$\mu_{i}^{\mathrm{theor}}=5\log_{10}D_{L} +25$,
where $D_L = cr(1+z)$ is the luminosity distance, and $r$ is 
given by (\ref{dis}).

We assume flat prior PDFs for the model parameters
over their entire assumed ranges.
The allowed range of $\Omega_{m}$ is the
same as the one used to constrain the $\phi$ function 
of eq. (\ref{cmbpeaks}) \cite{DL02}, i.e.  $\Omega_{m}\in[0.2, 0.6]$.
Finally, since we are interested in a wide range of models,
we choose the parameter $n$ to vary over a wide range $n\in[-4,4]$.

\begin{figure}
\includegraphics[scale=0.45]{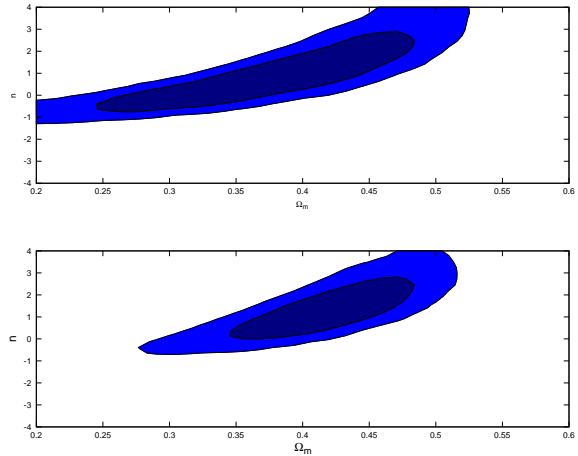}
\caption{Constraints from SN data (SNLS - upper panel, Union - lower panel).
$68\%$ and $95\%$ contours are presented.}
\label{fig1}
\end{figure}

The results are presented in Fig. \ref{fig1}.
As seen, the higher $\Omega_m$, the higher $n$ must be taken in order to obtain 
a satisfactory fit. Also, the Union data set puts tighter constraints on the
allowed range of the model parameters, although it prefers 
higher values of $\Omega_m$ and $n$. 

\subsubsection{CMB data}

\noindent
The second set of cosmological observations comprises measurement of the CMB angular power spectrum.
Here, we present 
the analysis of the CMB data which is restricted only to fitting 
the positions of the first two peaks and trough.
Such analysis therefore ignores the shape of the CMB power spectrum.
One, therefore. should keep in mind that full CMB constraints
are tighter\footnote{We justify this approach by emphasizing that backreaction effects are still not fully understood and there is an ongoing debate on
how large is the amplitude of these effects.
Analysis of the position of the CMB peaks is straightforward and it
quickly allows to test if models are consistent with the data.
Once the backreactions effects are better understood a full
analysis will be required.}.

If it is assumed that the early Universe 
(before and up to the last scattering instant) is well described by the FLRW model
then the CMB power spectrum can be parametrized by \cite{DL02}

\begin{equation}
l_m = l_a (m - \phi_m),
\label{cmbpeaks}
\end{equation}
where $l_1, l_2,$ and $l_3$ are the positions of the first, second and third peaks, and

\begin{equation}
l_a = \pi \frac{r_{*}}{r_s},
\end{equation}
where $r_{*}$ is the co-moving distance to the last scattering surface and $r_s$ is the size of the sound horizon at the last scattering instant. 
The function $\phi_m$ describes the phase shift of the $m$-th peak and is
mainly sensitive to pre-recombination physics. It depends on the baryon density ($\Omega_{b} h^2$), where $h=H_0/(100 $ Mpc km s$^{-1})$, on the ratio of the radiation to matter density at last scattering $[\rho_r(z_{*})/\rho_m(z_{*})=0.042(\Omega_m h^2)^{-1}(z_{*}/10^3)]$, where $z_{*}$ is the recombination redshift, on the spectral index ($n_s$), and on the density of the dark energy before recombination. 

We fit the positions of the first and second peaks, and the first trough 
of the CMB power spectrum \cite{WMAP3}. We assume that $n_s=1$ and neglect the density of dark energy before recombination. $r_{*}=c/H_0r(z_{*})$, $r(z)$ is given by (\ref{dis}) and $r_s=c(2/3k_{eq})\sqrt{6/R_{eq}}\ln[(\sqrt{1+R_{*}}+\sqrt{R_{*}+R_{eq}})/(1+\sqrt{R_{eq}})]$, where $R_{eq}=R/(1+z_{eq})$, $R_{*}=R/(1+z_{*})$, $k_{eq}=H_0\sqrt{2\Omega_{m} z_{eq}}$, $R=31500\Omega_bh^2(T_{CMB}/2.7K)^{-4}$ and $z_{eq}=2.5 \times 10^{4}\Omega_{m}h^2(T_{CMB}/2.7K)^{-4}$. We take $T_{CMB}=2.728$ and 
employ the fitting formulae for $z_{*}$ 
as provided in Ref. \cite{Hu96}. 

We use the position of the peaks and trough locations in the CMB power spectrum coming from the WMAP3 measurements \cite{WMAP3}. Those values were obtained in an 
almost model independent manner. The power spectra was fitted by the model composed of a Gaussian peak, parabolic trough and a second parabolic peak. Such model contains eight independent parameters, including peaks and though positions. Constraints were obtained using Markov Chain Monte Carlo method. The values and uncertainties of these parameters are the maximum and the width of the posterior probability distribution function. Additionaly the uncertainties include calibration uncertainty and cosmic variance. The likelihood function could be therefore written in the following form

\begin{eqnarray}
&& L_{CMB}(\Omega_m, n, \Omega_b h^2, h) \propto \nonumber \\
&& \exp \left[ -\frac{1}{2} \left( \left(\frac{l_1-220.8}{0.7}\right)^2+ \left(\frac{l_{3/2}-412.4}{1.9}\right)^2+ \right. \right. \nonumber \\ 
&& \left. \left. 
\left(\frac{l_2-530.9}{3.8}\right)^2 \right)\right].
\end{eqnarray}

We assume flat prior PDFs for the model parameters in the ranges (case 1): $\Omega_m \in [0.2, 0.6]$, $n \in [-4, 4]$, $h \in [0.64, 0.80]$ (using information from HST \cite{Freedman01}), $\Omega_b h^2 \in [0.0203, 0.0223]$ (using information from BBN \cite{Pettini08}).   

\begin{figure}
\includegraphics[scale=0.57]{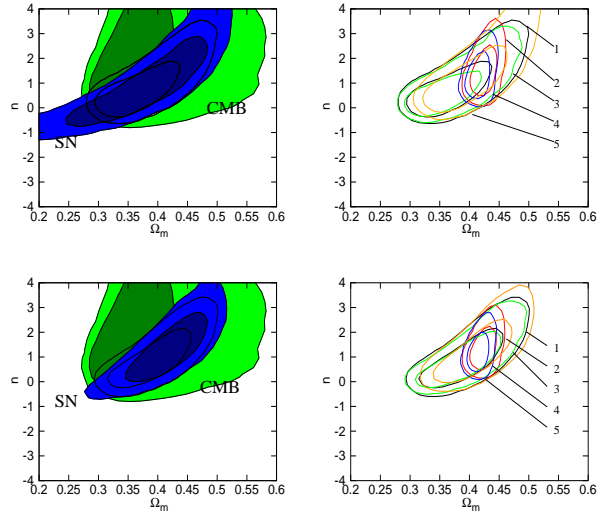}
\caption{Left panels: constraints from SN (blue), CMB (green) and SN+CMB (black) data (SNLS+CMB - upper panel, Union+CMB - lower panel). Right panels: comparison of constraints from SN+CMB data (SNLS+CMB - upper panel, Union+CMB - lower panel) with different prior assumptions on the parameters $h$ and $\Omega_b h^2$: case 1 (black), case 2 (red), case 3 (green), case 4 (blue), case 5 (orange).}
\label{fig2}
\end{figure}

The $68\%$ and $95\%$ contours of the posterior PDFs (after marginalization over $h$ and $\Omega_b h^2$),
obtained in the analysis with the CMB data as well as in the analysis with the joint data set (SNIa+CMB) (here the likelihood function has the following form $L=L_{SN}L_{CMB}$) are presented in Fig. \ref{fig2} (left panels). As seen both data sets prefer $n>-1$. However,  
unlike SNIa, CMB data prefers lower values of $\Omega_m$,
but still the best fits are consistent with each other.

\begin{figure}
\includegraphics[scale=0.57]{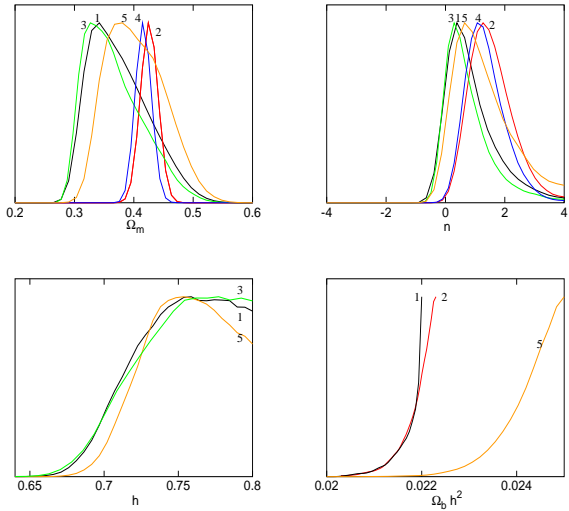}
\caption{Marginal posterior PDFs for model parameters (SNLS+CMB) with different prior assumption on $h$ and $\Omega_b h^2$ parameters: case 1 (black), case 2 (red), case 3 (green), case 4 (blue), case 5 (orange).}
\label{fig3a}
\end{figure}

\begin{figure}
\includegraphics[scale=0.57]{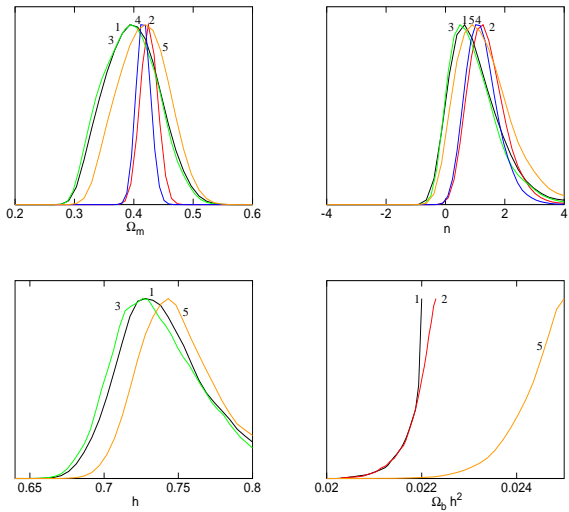}
\caption{Marginal posterior PDFs for model parameters (Union+CMB) with different prior assumption on $h$ and $\Omega_b h^2$ parameters: case 1 (black), case 2 (red), case 3 (green), case 4 (blue), case 5 (orange).}
\label{fig3b}
\end{figure}

The marginal posterior PDFs are presented in Figs.~\ref{fig3a} and \ref{fig3b} (black lines).
As seen, preferred values of $\Omega_m$ are higher than inferred
within the standard cosmological model. 
The means together with the $68\%$ Bayesian confidence intervals (credible intervals) and the posterior modes are presented in Table \ref{Tab1} and Table \ref{Tab2} for SNLS+CMB and Union+CMB respectively.
An interesting observation is that the mean of $n$ is around $1$. 
When $n=0$, the $H(z)$ does not depend on the spatial curvature.
Hence, if only expansion history is considered, in such a case the backreaction
effect  behaves as the cosmological constant. When 
$n=2$, the $H(z)$ depends on the spatial curvature as in FLRW models.

It is interesting to see how constraints on the $\Omega_m$ and $n$ are changed after changing the prior information on the $h$ and $\Omega_b h^2$ parameters. We consider four different situations: case 2 -- we fix the value of the Hubble parameter to the best fit value obtained in \cite{Freedman01}, i.e. $h=0.72$; case 3 -- we fix the value of $\Omega_b h^2$ to the best fit value obtained in \cite{Pettini08}, i.e. $\Omega_b h^2=0.0213$; case 4 -- we fix the values of both parameters ($h=0.72$ and $\Omega_b h^2=0.0213$). 
We observe that, CMB and SNIa data considered together prefer larger values of $\Omega_b h^2$ than expected from BBN constraints. Because of this we additionally check how the results are changed after allowing the $\Omega_b h^2$ parameter to take larger values, i.e. expand the prior range to $(\Omega_b h^2)^* \in  [0.020,0.025]$ (case 5). Constraints from SNIa and CMB data on the $\Omega_m$ and $n$ parameters for the cases described above are shown in the right panel of Figure \ref{fig2}. The largest changes occur when the Hubble parameter is fixed (blue and red contours). The values of $\Omega_m$ and $n$ are shifted to larger values and the constraints become tighter (especially for the matter density). On the other hand, fixing the value of $\Omega_b h^2$ does not give substantial changes in the constraints (green and blue contours). We can see that expanding the allowed range shifts the best 
fit values of $\Omega_m$ and $n$ upwards but does not improve the constraints (orange contours). This can also be seen on the marginal posterior PDF plots. One can additionally conclude that fixing the value of $\Omega_b h^2$ does not have any influence on the marginal posterior PDF for $h$ (green lines), but expanding the allowed range of $\Omega_b h^2$ changes it slightly (orange lines). Fixing the value of $h$ does not influence the posterior PDF of $\Omega_b h^2$ (red lines). SNIa and CMB data considered together prefer larger values of $\Omega_b h^2$, even when an extension of the allowed range is considered (orange lines). As one can see, changes in the constraints on the $\Omega_m$ and $n$ parameters are more prominent when the SNLS data set is considered. The means together with the $68\%$ Bayesian confidence intervals (credible intervals), and the posterior modes are presented in Table \ref{Tab1} and Table \ref{Tab2} for SNLS+CMB and Union+CMB respectively. 

\subsubsection{BAO data}

\begin{figure}
\includegraphics[scale=0.57]{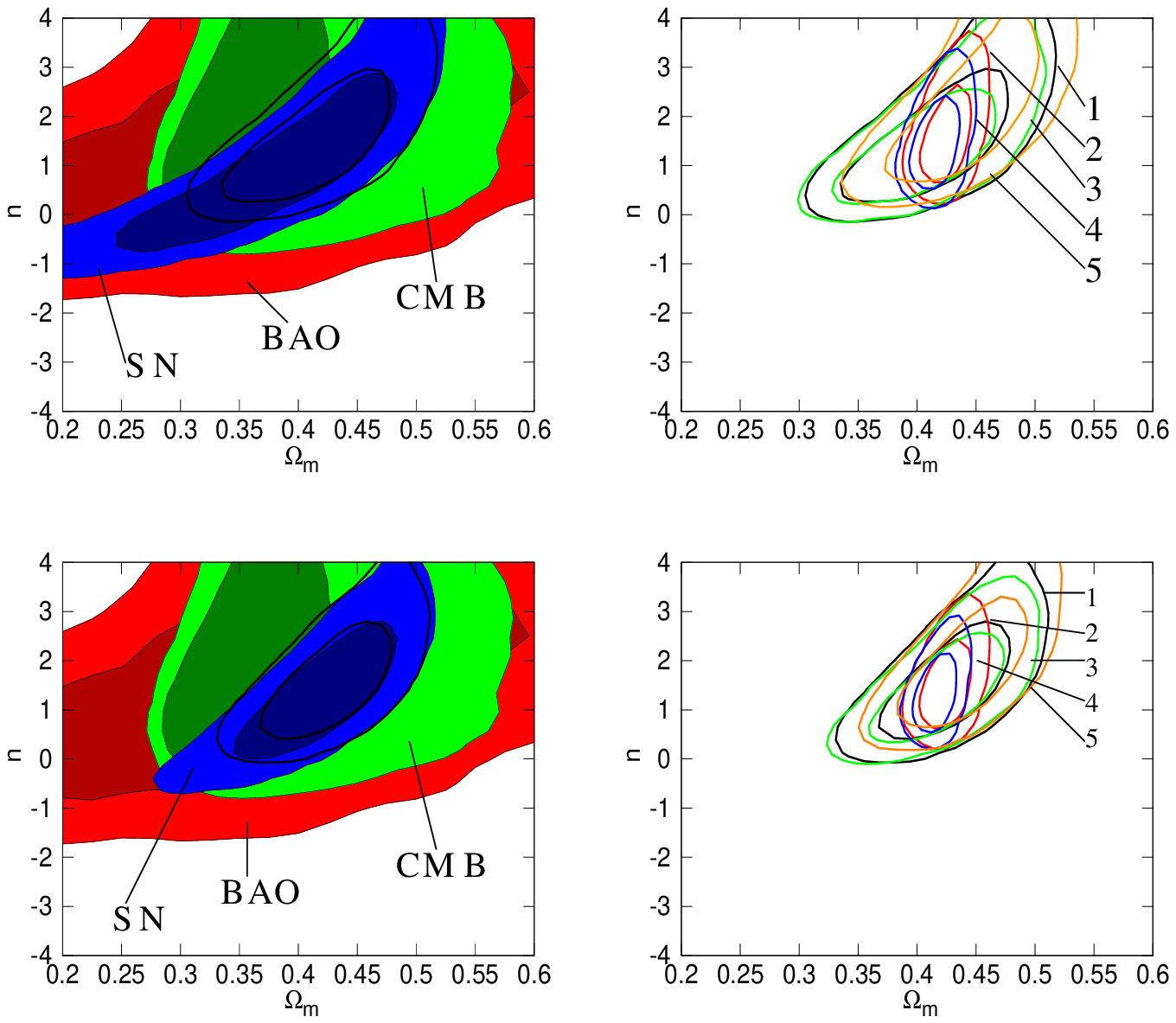}
\caption{Left panels: constraints from SNIa (blue), CMB (green), BAO (red) and SNIa+CMB+BAO (black) data sets (SNLS+CMB+BAO - upper panel, Union+CMB+BAO - lower panel). Right panels: comparison of constrains from SN+CMB+BAO data (SNLS+CMB+BAO - upper panel, Union+CMB+BAO - lower panel) with different prior assumption on $h$ and $\Omega_b h^2$ parameters: case 1 (black), case 2 (red), case 3 (green), case 4 (blue), case 5 (orange).}
\label{fig4}
\end{figure}

\noindent
In addition to the geometric measurements described above, 
we study constraints obtained from 
the measured dilation scale of the BAO in the redshift space power-spectrum of 46,748 luminous red galaxies (LRG) from the Sloan Digital Sky Survey
(SDSS). The dilation scale is defined as
\begin{equation}
D_V = \left[ D_A^2 \frac{cz}{H(z)} \right]^{1/3},
\label{dvdf}
\end{equation}
where $D_A$ is the co-moving angular diameter distance and $H(z)$ is the Hubble parameter as afunction of redshift. The measured value of the dilation scale at $z=0.35$ 
is 1370 $\pm$ 64 Mpc. 
It should be noted that the value of 1370 $\pm$ 64 Mpc was obtained
within the framework of linear perturbations imposed on the homogeneous FLRW background. Instead,
such an analysis should be carried out within the framework of the model considered in this paper. Otherwise, we 
should be aware of possible systematical errors.
When the geometry of the spacetime is not FLRW the possible sources of errors are:
(1) the sound horizon can be
distorted and can be of a different size in parallel 
and perpendicular directions; (2) the expansion rate 
can be different with respect to 
parallel and perpendicular directions;
(3) the redshift distortions, if analysed within the inhomogeneous
model, might lead to estimates different from those received within the standard approach; (4) 
another source of error comes from the fact that 
in their analysis Eisenstein et al. converted the redshifts of
LRG galaxies to distances assuming the $\Lambda$CDM model.
Despite these uncertainties we proceed with the analysis 
to see how the BAO data can possibly constrain the data.
As seen from Fig. \ref{fig4}, the measurements of the
dilation scale at $z=0.35$ do not put tight constraints
on the parameters of the model.
 At higher redshifts
the constraints will be tighter, and thus
an analysis of the BAO within the backreaction model will be required
when future observational data is available.
The likelihood function for the BAO has the following form
\begin{equation}
L_{BAO}(\Omega_m, n, h) \propto \exp \left[- \frac {(D_V^{\mathrm{theor}}-D_V^{\mathrm{obs}})^2}{2\sigma_{BAO}^2} \right].
\end{equation}
We assume flat prior PDFs for the parameters within the ranges described above for case 1.
The $68\%$ and $95\%$ contours of the posterior PDF (marginalized over $h$ and $\Omega_b h^2$ ) for the joint constraints from the supernovae,
CMB and BAO data (with the likelihood function of the following form $L=L_{SN} L_{CMB} L_{BAO}$) are presented in Fig. \ref{fig4} (black lines). 

\begin{figure}
\includegraphics[scale=0.57]{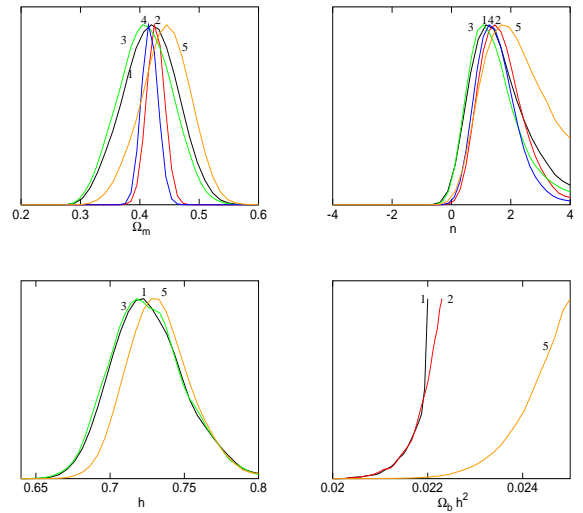}
\caption{Marginal posterior PDFs for model parameters (SNLS+CMB+BAO) with different prior assumptions on the parameters $h$ and $\Omega_b h^2$: case 1 (black), case 2 (red), case 3 (green), case 4 (blue), case 5 (orange).}
\label{fig5a}
\end{figure}

\begin{figure}
\includegraphics[scale=0.57]{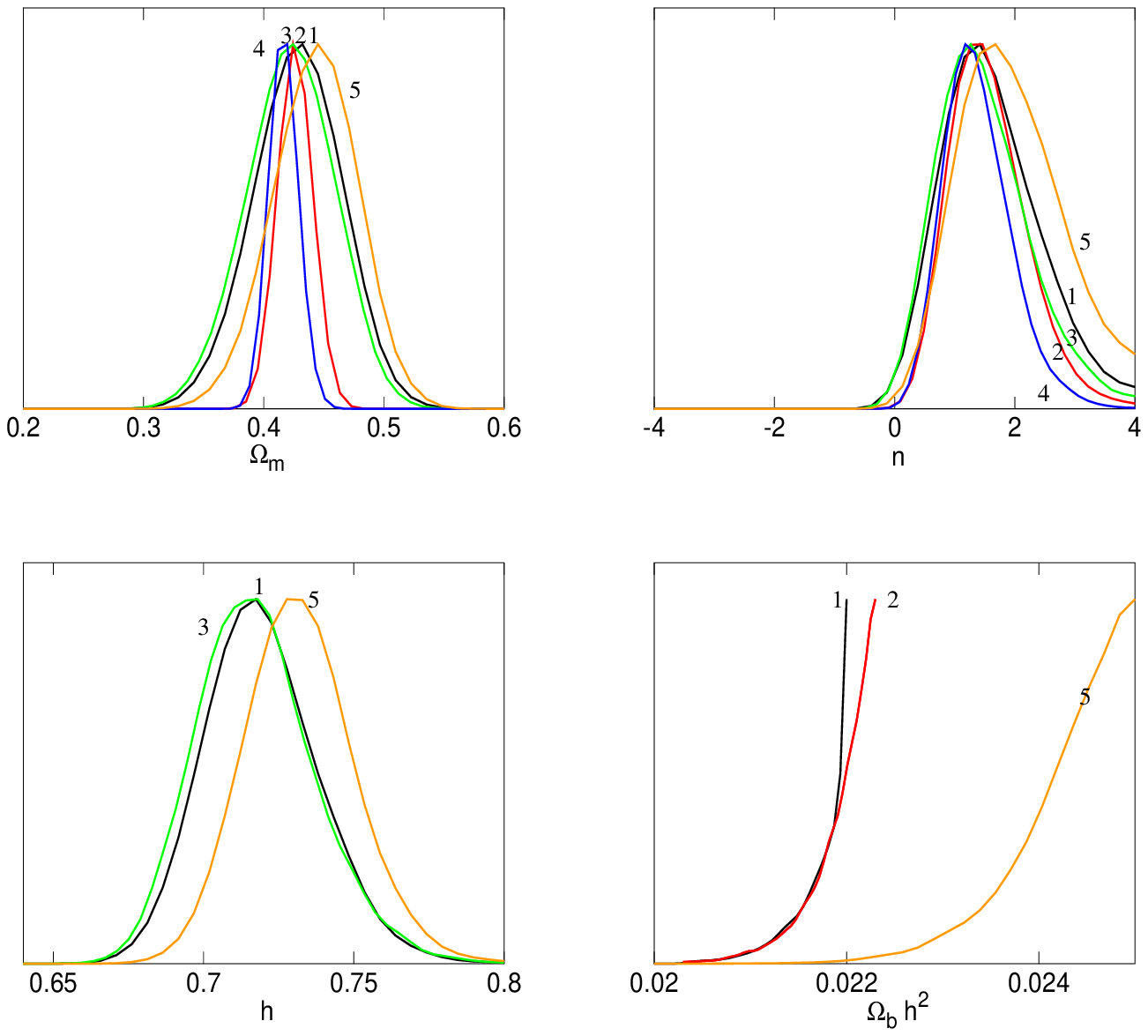}
\caption{Marginal posterior PDFs for model parameters (Union+CMB+BAO) with different prior assumptions on the parameters $h$ and $\Omega_b h^2$: case 1 (black), case 2 (red), case 3 (green), case 4 (blue), case 5 (orange).}
\label{fig5b}
\end{figure}

The marginal posterior PDFs are presented in Figs.~\ref{fig5a} and \ref{fig5b} (black lines). 
In comparison with the preceding section, the PDFs are similar
except for $n$, which is now shifted to higher values.
The means of the marginal posterior PDFs
together with the $68\%$ credible intervals are presented in Table \ref{Tab1} and Table \ref{Tab2} for SNLS+CMB+BAO and Union+CMB+BAO respectively.
For example, $n$ based on Union+CMB+BAO is now 
$1.60^{+0.80}_{-0.76}$ in comparison
to $1.02^{+0.79}_{-0.77}$ inferred only from  Union+CMB.
We will come back to the relation between
the curvature and backreaction  effects in Sec. \ref{curidx}.
We also check how the results are sensitive to changes in the prior information for the $h$ and $\Omega_b h^2$ parameters. We consider all cases described above. Observe that changes are most prominent when the value of $h$ is fixed. The contours become narrower (red and blue contours on the right panels of Figure \ref{fig4}), which is due to tigher constraints on the $\Omega_m$ parameter. The values of  $\Omega_m$ and $n$ do not change significantly. When we expand the allowed prior range for the parameter $\Omega_b h^2$, the values of $\Omega_m$ and $n$ are slightly shifted upwards, and the constraints on $n$ become weaker (see also the posterior PDFs presented in Figure \ref{fig5a} and Figure \ref{fig5b}). Constraints on $h$ are also changed in this case: its value is shifted upwards. 

An interesting point is that the PDF of $\Omega_b h^2$ increases
with $\Omega_b h^2$ and does not seem to decrease in the considered range.
This implies that the maximum of the PDF is out of the range and might suggest 
a tension between SNIa+CMB, SNIa+CMB+BAO and BBN constraints on $\Omega_b h^2$.

\begin{table}
\caption{\label{Tab1} Means of the marginal posterior PDFs for the model parameters, together with the $68\%$ credible intervals for the SNIa+CMB and SNIa+CMB+BAO data sets. The corresponding values for the posterior modes are presented in brackets. }
\begin{ruledtabular}
\begin{tabular}{lcc}
& SNLS+CMB  & SNLS+CMB+BAO  \\ \hline
$\Omega_m$&$0.37^{+0.06}_{-0.04} (0.32)$&$0.42^{+0.04}_{-0.05} (0.40)$\\
$n$ &$0.87^{+0.78}_{-0.71} (0.00)$&$1.59^{+0.91}_{-0.85} (1.48)$\\
$h$ &$0.75^{+0.03}_{-0.03} (0.79)$  &$0.73^{+0.02}_{-0.03} (0.73)$  \\
$\Omega_b h^2$& $0.0219^{+0.0003}_{-0.0003} (0.0223)$&$0.0219^{+0.0003}_{-0.0003} (0.0223)$  \\
\hline
$\Omega_m$&$0.43^{+0.01}_{-0.02} (0.43)$&$0.43^{+0.01}_{-0.02} (0.43)$\\
$n$ &$1.55^{+0.69}_{-0.69} (1.48)$&$1.70^{+0.71}_{-0.70} (1.68)$\\
$h$ &$0.72$  &$0.72$  \\
$\Omega_b h^2$& $0.0219^{+0.0003}_{-0.0003} (0.0222)$&$0.0219^{+0.0003}_{-0.0003} (0.0223)$ \\
\hline
$\Omega_m$&$0.37^{+0.05}_{-0.05} (0.31)$&$0.41^{+0.04}_{-0.05} (0.40)$\\
$n$ &$0.76^{+0.67}_{-0.66} (-0.05)$&$1.46^{+0.83}_{-0.79} (1.17)$\\
$h$ &$0.75^{+0.03}_{-0.03} (0.80)$  &$0.72^{+0.03}_{-0.02} (0.73)$  \\
$\Omega_b h^2$& $0.0213$&$0.0213$  \\
\hline
$\Omega_m$&$0.42^{+0.01}_{-0.01} (0.41)$&$0.42^{+0.01}_{-0.01} (0.41)$\\
$n$ &$1.36^{+0.64}_{-0.61} (1.15)$&$1.53^{+0.62}_{-0.63} (1.29)$\\
$h$ &$0.72$  &$0.72$  \\
$\Omega_b h^2$& $0.0213$&$0.0213$  \\
\hline
$\Omega_m$&$0.40^{+0.05}_{-0.05} (0.34)$&$0.44^{+0.04}_{-0.04} (0.45)$\\
$n$ &$1.27^{+0.97}_{-0.89} (0.27)$&$2.01^{+1.10}_{-0.88} (2.00)$\\
$h$ &$0.75^{+0.03}_{-0.02} (0.79)$  &$0.73^{+0.02}_{-0.02} (0.73)$  \\
$(\Omega_b h^2)^*$& $0.0243^{+0.0005}_{-0.0005} (0.0248)$&$0.0243^{+0.0005}_{-0.0006} (0.0245)$  \\
\end{tabular}
\end{ruledtabular}
\end{table}

\begin{table}
\caption{\label{Tab2} Means of the marginal posterior PDFs for the model parameters together with the $68\%$ credible intervals for the SNIa+CMB and SNIa+CMB+BAO data sets. The corresponding values for the posterior modes are presented in brackets.}
\begin{ruledtabular}
\begin{tabular}{lcc} 
& Union+CMB & Union+CMB+BAO \\ \hline
$\Omega_m$ & $0.40^{+0.04}_{-0.05} (0.36)$ & $0.43^{+0.03}_{-0.04} (0.42)$ \\
$n$ & $1.02^{+0.79}_{-0.77} (0.35)$ & $1.60^{+0.80}_{-0.76} (1.14)$\\
$h$ & $0.74^{+0.03}_{-0.03} (0.76)$ & $0.72^{+0.02}_{-0.02} (0.72)$ \\
$\Omega_b h^2$ & $0.0219^{+0.0003}_{-0.0003} (0.0223)$ & $0.0219^{+0.0003}_{-0.0003} (0.0223)$  \\
\hline
$\Omega_m$ & $0.42^{+0.02}_{-0.01} (0.43)$ & $0.43^{+0.01}_{-0.02} (0.42)$ \\
$n$ & $1.41^{+0.58}_{-0.58} (1.51)$ & $1.57^{+0.61}_{-0.60} (1.34)$\\
$h$ & $0.72$ & $0.72$ \\
$\Omega_b h^2$ & $0.0219^{+0.0003}_{-0.0003} (0.0223)$ & $0.0219^{+0.0003}_{-0.0003} (0.0223)$  \\
\hline
$\Omega_m$ & $0.39^{+0.05}_{-0.04} (0.38)$ & $0.42^{+0.04}_{-0.03} (0.42)$ \\
$n$ & $0.97^{+0.76}_{-0.75} (0.48)$ & $1.48^{+0.73}_{-0.74} (1.40)$\\
$h$ & $0.73^{+0.03}_{-0.02} (0.74)$ & $0.72^{+0.02}_{-0.02} (0.72)$ \\
$\Omega_b h^2$ & $0.0213$ & $0.0213$  \\
\hline
$\Omega_m$ & $0.42^{+0.01}_{-0.01} (0.41)$ & $0.42^{+0.01}_{-0.01} (0.41)$ \\
$n$ & $1.25^{+0.53}_{-0.52} (1.07)$ & $1.39^{+0.53}_{-0.53} (1.23)$\\
$h$ & $0.72$ & $0.72$ \\
$\Omega_b h^2$ & $0.0213$ & $0.0213$  \\
\hline
$\Omega_m$ & $0.41^{+0.05}_{-0.04} (0.39)$ & $0.44^{+0.03}_{-0.03} (0.44)$ \\
$n$ & $1.30^{+0.85}_{-0.84} (0.83)$ & $1.92^{+0.84}_{-0.81} (1.94)$\\
$h$ & $0.75^{+0.02}_{-0.03} (0.77)$ & $0.73^{+0.02}_{-0.02} (0.74)$ \\
$(\Omega_b h^2)^*$ & $0.0244^{+0.0005}_{-0.0006} (0.0249)$ & $0.0243^{+0.0005}_{-0.0005} (0.0249)$  \\
\end{tabular}
\end{ruledtabular}
\end{table}

\subsection{Models comparison}

\subsubsection{${\Lambda}$CDM vs the backreaction model}

\begin{figure}
\includegraphics[scale=0.45]{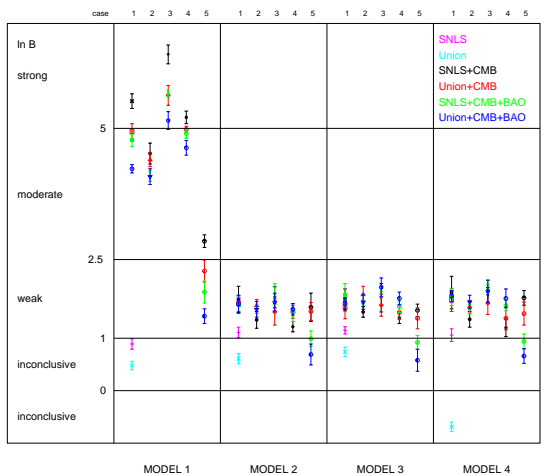}
\caption{Values of the logarithm of the Bayes factor $\ln B_{0i}$ calculated for the $\Lambda$CDM model and the model indexed by $i$ for five different prior cases and different data sets (SNLS - magenta, Union - cyan, SNLS+CMB - black, Union+CMB - red, SNLS+CMB+BAO - green, Union+CMB+BAO - blue).}
\label{fig6}
\end{figure}

\noindent
In this section we present the comparison between the model considered 
above (which will be referred to as model 1) and the $\Lambda$CDM model (which will be referred to as model 0). 
In the  Bayesian framework, 
models are compared not only by 
how well they fit the data,
but also by their complexity (see \cite{RT08} for a review).
The best model from the set of models under consideration is the one with the greatest value of the probability in the light of data defined as
\begin{equation}
 P(M_{i}|D)=\frac{P(D|M_{i})P(M_{i})}{P(D)},
\end{equation}
where $M_i$  is a model under consideration and $P(M_{i})$ is the prior probability of a model. If we have no foundation 
for favouring one model in the set of models under consideration
over another,
 we usually assume the same value of the prior quantity for all models, i.e. $P(M_{i})=1/K$, where $K$ is the number of models. $P(D)$ is the normalization constant. $P(D|M_{i})$ is called the marginal likelihood (or the evidence) and has the following form
\begin{equation}
P(D|M_{i})=\int L(\bar{\theta})P(\bar{\theta}|M_{i}) {\rm d} \bar{\theta} \equiv E_{i}.
\label{evidence}
\end{equation} 
It is convenient to consider the ratio of posterior probabilities for the models which we want to compare. If prior probabilities for those models are equal, 
then the posterior ratio reduces
to the ratio of the evidences. This ratio is called the Bayes factor ($B_{ij}\equiv E_{i}/E_{j})$. The values of $B_{ij}$ are interpreted as follows: $0<\ln B_{ij}<1$ as inconclusive,   
$1<\ln B_{ij}<2.5$ as weak, $2.5<\ln B_{ij}<5$ as moderate and $\ln B_{ij}>5$ as strong evidence in favour of a model indexed by $i$ with respect to a model indexed by $j$. The evidence [eq. (\ref{evidence})] was calculated using the CosmoNest code. We generated five chains for each case to obtain the uncertainty in the computed value of the evidence.  We 
performed our calculations 
using the different prior assumptions for the parameters $h$ and $\Omega_b h^2$, which were described in the previous section. 
The values of the logarithm of the Bayes factor calculated 
for the $\Lambda$CDM model (model 0) vs 
model 1 (i.e. the model presented in Sec. \ref{parest}) -- $B_{01}$  -- 
are presented in Table \ref{Tab3} and in Figure \ref{fig6}.

\begin{table*}
\caption{\label{Tab3} Values of the logarithm of the Bayes factor $\ln B_{ij}$ calculated for the model indexed by $i$ and the model indexed by $j$ for different priors and different data sets.} 
\begin{ruledtabular}
\begin{tabular}{llccccccc}
Case &Data set & $\ln B_{01}$ & $\ln B_{21}$ & $\ln B_{02}$ & $\ln B_{03}$ & $\ln B_{31}$ & $\ln B_{04}$ & $\ln B_{41}$\\ \hline
 1&SNLS & $0.89 \pm 0.10$ & $-0.22 \pm 0.10$ & $1.11 \pm 0.10$ & $1.15 \pm 0.07$ & $-0.25 \pm 0.07$ & $1.06 \pm 0.12$& $-0.17 \pm 0.12$\\
  &Union & $0.48 \pm 0.08$ & $-0.13 \pm 0.09$ & $0.61 \pm 0.10$ & $0.74 \pm 0.09$ & $-0.26 \pm 0.08$ & $-0.69 \pm 0.09$& $-0.21 \pm 0.08$\\
  &SNLS+CMB & $5.52 \pm 0.14$ & $3.78 \pm 0.27$ & $1.74 \pm 0.25$ & $1.73 \pm 0.21$ & $3.79 \pm 0.24$ & $1.87 \pm 0.31$& $3.65 \pm 0.33$ \\
  &Union+CMB & $4.94 \pm 0.15$ & $3.28 \pm 0.13$ & $1.66 \pm 0.15$ & $1.59 \pm 0.22$ & $3.34 \pm 0.21$ & $1.73 \pm 0.22$& $3.21 \pm 0.21$\\
  &SNLS+CMB+BAO & $4.77 \pm 0.12$ & $3.13 \pm 0.19$ & $1.64 \pm 0.17$ & $1.83 \pm 0.21$ & $2.93 \pm 0.23$ &$1.78 \pm 0.16$ & $2.99 \pm 0.18$ \\
  &Union+CMB+BAO & $4.23 \pm 0.08$ & $2.58 \pm 0.18$ & $1.65 \pm 0.17$ & $1.66 \pm 0.11$ & $2.57 \pm 0.12$ & $1.80 \pm 0.11$& $2.43 \pm 0.12$\\
\hline
 2 &SNLS+CMB & $4.52 \pm 0.20$ & $3.17 \pm 0.24$ & $1.35 \pm 0.16$ & $1.50 \pm 0.10$ & $3.02 \pm 0.21$ & $1.36 \pm 0.15$& $3.16 \pm 0.23$ \\
  &Union+CMB & $4.40 \pm 0.13$ & $2.78 \pm 0.10$ & $1.62 \pm 0.12$ & $1.83 \pm 0.16$ & $2.57 \pm 0.15$ & $1.60 \pm 0.12$& $2.80 \pm 0.10$\\
  &SNLS+CMB+BAO & $4.09 \pm 0.10$ & $2.53 \pm 0.09$ & $1.56 \pm 0.09$ & $1.73 \pm 0.10$ & $2.36 \pm 0.10$ &$1.55 \pm 0.09$ & $2.54 \pm 0.09$ \\
  &Union+CMB+BAO & $4.08 \pm 0.15$ & $2.52 \pm 0.12$ & $1.55 \pm 0.15$ & $1.69 \pm 0.14$ & $2.38 \pm 0.12$ & $1.69 \pm 0.13$& $2.38 \pm 0.10$\\
\hline
3 &SNLS+CMB & $6.41 \pm 0.18$ & $4.62 \pm 0.17$ & $1.78 \pm 0.20$ & $1.77 \pm 0.27$ & $4.64 \pm 0.24$ & $1.82 \pm 0.15$& $4.59 \pm 0.11$ \\
  &Union+CMB & $5.63 \pm 0.19$ & $4.13 \pm 0.23$ & $1.50 \pm 0.25$ & $1.63 \pm 0.21$ & $4.00 \pm 0.19$ & $1.67 \pm 0.22$& $3.96 \pm 0.20$\\
  &SNLS+CMB+BAO & $5.67 \pm 0.06$ & $3.82 \pm 0.20$ & $1.84 \pm 0.20$ & $1.89 \pm 0.11$ & $3.77 \pm 0.12$ &$2.02 \pm 0.08$ & $3.65 \pm 0.09$ \\
  &Union+CMB+BAO & $5.15 \pm 0.17$ & $3.47 \pm 0.13$ & $1.69 \pm 0.17$ & $1.97 \pm 0.18$ & $3.18 \pm 0.14$ & $1.90 \pm 0.21$& $3.25 \pm 0.18$\\
\hline
4 &SNLS+CMB & $5.21 \pm 0.12$ & $3.99 \pm 0.13$ & $1.22 \pm 0.10$ & $1.39 \pm 0.11$ & $3.82 \pm 0.13$ & $1.20 \pm 0.17$& $4.01 \pm 0.19$ \\
  &Union+CMB & $4.98 \pm 0.06$ & $3.48 \pm 0.12$ & $1.50 \pm 0.12$ & $1.49 \pm 0.13$ & $3.49 \pm 0.13$ & $1.38 \pm 0.22$& $3.59 \pm 0.22$\\
  &SNLS+CMB+BAO & $4.90 \pm 0.09$ & $3.41 \pm 0.16$ & $1.49 \pm 0.17$ & $1.60 \pm 0.11$ & $3.30 \pm 0.09$ &$1.62 \pm 0.10$ & $3.28 \pm 0.08$ \\
  &Union+CMB+BAO & $4.63 \pm 0.14$ & $3.08 \pm 0.15$ & $1.55 \pm 0.11$ & $1.76 \pm 0.12$ & $2.87 \pm 0.16$ & $1.76 \pm 0.18$& $2.88 \pm 0.21$\\
\hline
5 &SNLS+CMB & $2.85 \pm 0.12$ & $1.26 \pm 0.27$ & $1.59 \pm 0.27$ & $1.53 \pm 0.12$ & $1.32 \pm 0.12$ & $1.77 \pm 0.14$& $1.08 \pm 0.14$ \\
  &Union+CMB & $2.28 \pm 0.21$ & $0.77 \pm 0.20$ & $1.51 \pm 0.17$ & $1.38 \pm 0.20$ & $0.90 \pm 0.22$ & $1.47 \pm 0.22$& $0.81 \pm 0.24$\\
  &SNLS+CMB+BAO & $1.88 \pm 0.21$ & $0.89 \pm 0.22$ & $0.99 \pm 0.15$ & $0.92 \pm 0.13$ & $0.96 \pm 0.20$ &$0.94 \pm 0.14$ & $0.93 \pm 0.21$ \\
  &Union+CMB+BAO & $1.42 \pm 0.14$ & $0.73 \pm 0.22$ & $0.69 \pm 0.20$ & $0.58 \pm 0.21$ & $0.83 \pm 0.23$ & $0.66 \pm 0.14$& $0.75 \pm 0.17$\\
\end{tabular}
\end{ruledtabular}
\end{table*}

The comparison in the light of the SNIa data does not give conclusive results
-- this data set has not enough information to favour one model over another. After the inclusion of information coming from the CMB, 
there is strong (SNLS+CMB) and almost strong (Union+CMB)
evidence in favour of the $\Lambda$CDM model over the inhomogeneous one.
When we include information coming from the BAO, the values of the Bayes factor become smaller in both cases and the evidence in favour the $\Lambda$CDM model is moderate.

Let us consider the influence of the prior information for the $h$ and $\Omega_b h^2$ parameters on the evidence. The most prominent change is when the range of $\Omega_b h^2$ is extended (case 5). The evidence in favour of the $\Lambda$CDM model is moderate for the SNLS+CMB data set and weak otherwise. When the value of $h$ is fixed (case 2), the value of $\ln B$ becomes smaller for all data sets and the evidence in favour of the $\Lambda$CDM model is moderate. On the other hand, fixing the value of $\Omega_b h^2$ (case 3) shifts the value of $\ln B$ upwards and the evidence in favour the $\Lambda$CDM model becomes strong. Fixing both $h$ and $\Omega_b h^2$ (case 4) does not change the final conclusions. 

\begin{figure}
\includegraphics[scale=0.45]{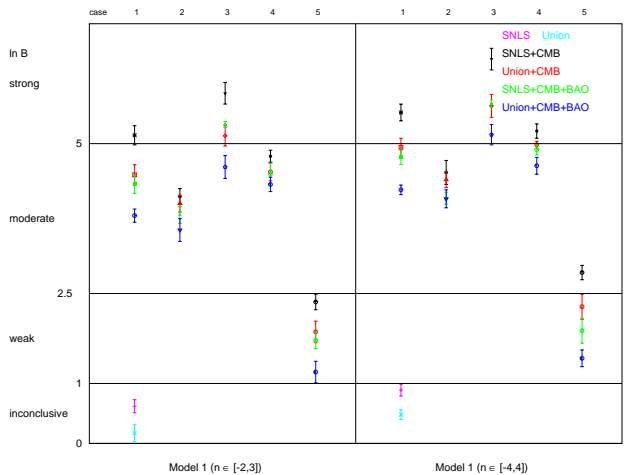}
\caption{Values of logarithm of the Bayes factor $\ln B_{01}$ calculated for the $\Lambda$CDM model and Model 1 (with two different prior assumptions on $n$ parameter) for five different prior cases (regarding $h$ and $\Omega_b h^2$ parameters) and different data sets (SNLS - magenta, Union - cyan, SNLS+CMB - black, Union+CMB - red, SNLS+CMB+BAO - green, Union+CMB+BAO - blue)}
\label{fig7}
\end{figure}

It is also interesting to see how strong is the influence of the prior information regarding the $n$ parameter on the final results. We repeat our calculations for the case in which the prior range for the $n$ parameter is changed, i.e. for $n \in [-2,3]$.
The values of the logarithm of the Bayes factor calculated with respect to the $\Lambda$CDM model for different data sets and different prior cases regarding the $h$ and $\Omega_b h^2$ parameters are presented in Table \ref{Tab6} and in Figure \ref{fig7}. 

\begin{table*}
\caption{\label{Tab6} Values of the logarithm of the Bayes factor $\ln B_{01}$ calculated for the $\Lambda$CDM model and model 1 with assumption that $n \in[-2,3]$ for different priors and different data sets.} 
\begin{ruledtabular}
\begin{tabular}{lccccc}
Data set & case 1 & case 2 & case 3 & case 4 & case 5 \\
\hline
SNLS &$0.62 \pm 0.11$&-- & -- &-- & --\\
Union &$0.17 \pm 0.14$& --& --&-- &--  \\
SNLS+CMB & $5.14 \pm 0.16$&$4.12 \pm 0.13$&$5.84 \pm 0.18$&$4.79 \pm 0.10$& $2.36 \pm 0.13$\\
Union+CMB &$4.48 \pm 0.17$&$4.00 \pm 0.14$&$5.13 \pm 0.17$ &$4.53 \pm 0.15$& $1.86 \pm 0.18$     \\
SNLS+CMB+BAO &$4.32 \pm 0.15$&$3.81 \pm 0.14$&$5.32 \pm 0.05$&$4.49 \pm 0.18$&$1.72 \pm 0.14$ \\
Union+CMB+BAO&$3.80 \pm 0.11$&$3.56 \pm 0.19$&$4.61 \pm 0.19$ &$4.32 \pm 0.12$& $1.19 \pm 0.18$     \\
\end{tabular}
\end{ruledtabular}
\end{table*}

Observe that, when the prior range for $n$ is restricted, the values of the logarithm of the Bayes factor become smaller, but the differences in 
the case in which $n \in [-4,4]$ are small. Considering the comparison with the $\Lambda$CDM model, one finds that in general the final conclusions do not change. 

\subsubsection{Relation between the average curvature and the curvature index}\label{curidx}

\noindent
In the preceding subsection we could see that the 
Bayesian method of model comparison prefers the $\Lambda$CDM model over the backreaction model.
We should be aware that the backreaction model -- model 1 -- 
is based on the assumptions 
(\ref{ansatz1}) and (\ref{ansatz2}).
If these assumptions are 
 changed, it is possible to obtain a model with a better fit.

Below we present models in which the assumption (\ref{ansatz2})
is replaced with:

\begin{itemize}
\item model 2

\begin{equation}
k(z) =0.
\end{equation}

We  emphasize that $k=0$ comes from a modification of the assumption
(\ref{ansatz2}) only. It is not the same as assuming  that
$\av{\mathcal{R}} =0$. As seen from (\ref{intcond})
the assumption of $\av{\mathcal{R}} =0$ leads to 
$\mathcal{Q} \sim a^{-6}$, which means that the backreaction
is strong in the early Universe and its value decreases
with time.

The results of the model comparison are presented in Table \ref{Tab3} and in Figure \ref{fig6}.
As can be seen, both for the SNIa+CMB and
SNIa+CMB+BAO data sets,
there is moderate evidence
to favour the backreaction model with $k=0$ over the 
model with relation (\ref{ansatz2}). 
Note that there is only weak evidence to favour the $\Lambda$CDM model over model 2. A change in the prior information regarding $h$ and $\Omega_b h^2$ does not change the final conclusions for all considered cases, besides case 5. When the range for the parameter $\Omega_b h^2$ is extended, the evidence against model 1 becomes smaller. 

\item model 3

\begin{equation}
k(z)=-\frac{1}{p} (1+z)^{-(n+2)}.
\label{pm3}
\end{equation}

We assume a flat prior PDF for the additional parameter in the range $p \in [0,100]$. 
Observe that, model 2 fits the observations better than model 1. If indeed $k =0$
is favoured, then we should obtain that the best model with $k$ given by (\ref{pm3})
is the one with $p=100$. However, this is not the case and 
as seen from Fig.~\ref{fig6a}, the marginal posterior PDF is almost flat over a very wide range --
there is only a little difference between $p=100$ and the best-fit value.  This conclusion is confirmed bu observing the value of the logarithm of the Bayes factor: there is not enough information to favour model 2 over model 3. 
The means and modes of the marginal posterior PDFs for the model parameters are presented in Table \ref{Tab4}. The values of the logarithm of the Bayes factor, calculated for the $\Lambda$CDM model and model 3 ($\ln B_{03}$), as well as for model 3 and model 1 ($\ln B_{31}$) are compared in Table \ref{Tab3} (see also Figure \ref{fig6}). 
One may conclude that there is weak evidence to favour the $\Lambda$CDM model over model 3 and moderate evidence in favour of model 3 over model 1 (for the SNIa+CMB and SNIa+CMB+BAO data sets). The conclusion is changed when the range for $\Omega_b h^2$ is extended: for the SNIa+CMB+BAO data set, the evidence in favour of the $\Lambda$CMD model becomes inconclusive.  

\item model 4

\begin{equation}
k(z) =-(1+z)^{-(n+2+m)}.
\end{equation}

We assume a flat prior PDF for the additional parameter in the range $m \in [0,5]$.
The marginal posterior PDF for the parameter $m$ is presented in Fig. \ref{fig6b}.
The means and modes of the marginal posterior PDF for the model parameters are presented in Table \ref{Tab5}. The values of logarithm of the Bayes factor calculated for the $\Lambda$CDM model and model 4 ($\ln B_{04}$) as well as for model 4 and model 1 ($\ln B_{41}$) are compared in Table \ref{Tab3} (see also Figure \ref{fig6}).

\begin{figure}
\includegraphics[scale=0.57]{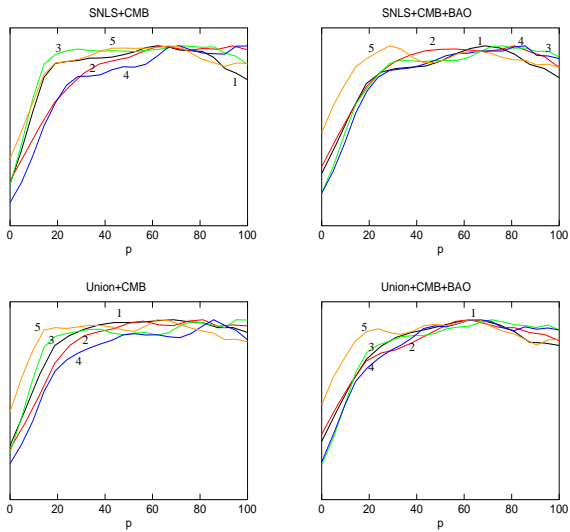}
\caption{Marginal posterior PDF for the parameter $p$ of model 3 with different prior assumptions on the $h$ and $\Omega_b h^2$ parameters: case 1 (black), case 2 (red), case 3 (green), case 4 (blue), case 5 (orange). 
}
\label{fig6a}
\end{figure}

\begin{figure}
\includegraphics[scale=0.57]{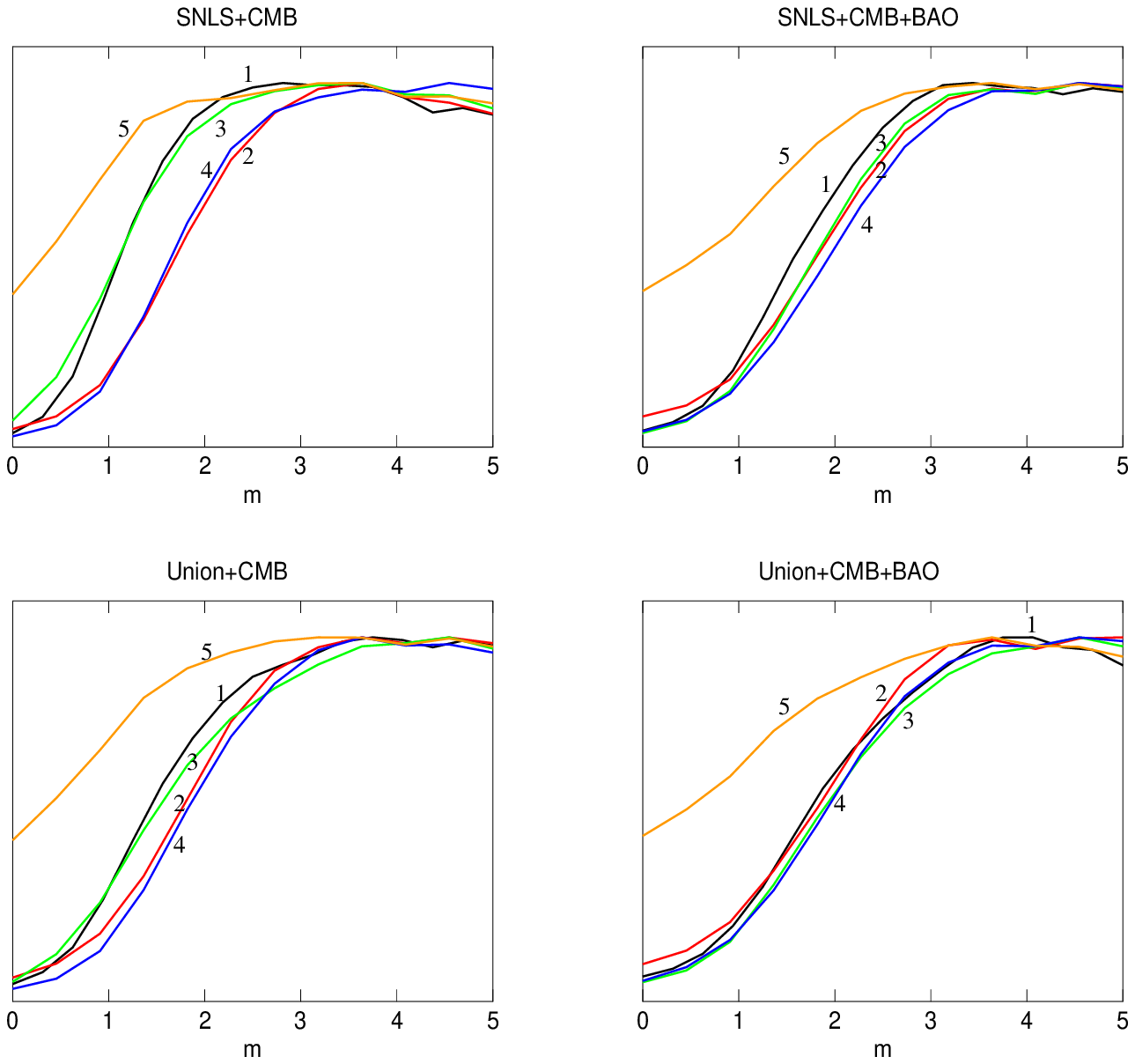}
\caption{Marginal posterior PDF for the parameter $m$ of model 4 with different prior assumptions on the $h$ and $\Omega_b h^2$ parameters: case 1 (black), case 2 (red), case 3 (green), case 4 (blue), case 5 (orange).}
\label{fig6b}
\end{figure}

\begin{table*}
\caption{\label{Tab4}Mean of the marginal posterior PDF for the parameters of model 3, together with the $68\%$ credible interval, for the SNIa+CMB and SNIa+CMB+BAO data sets. The corresponding values of posterior mode are presented in brackets. }
\begin{ruledtabular}
\begin{tabular}{lcccc}
& SNLS+CMB & SNLS+CMB+BAO & Union+CMB & Union+CMB+BAO \\  \hline
$\Omega_m$ & $0.24^{+0.04}_{-0.03}(0.20)$ & $0.28^{+0.06}_{-0.05} (0.27)$ & $0.26^{+0.05}_{-0.05} (0.20)$ & $0.31^{+0.06}_{-0.07} (0.31)$ \\
$n$ & $-0.18^{+0.28}_{-0.33}(-0.37)$ & $0.10^{+0.54}_{-0.55} (-0.20)$ &$-0.13^{+0.37}_{-0.41} (-0.53)$ & $0.27^{+0.70}_{-0.67}(0.16)$\\
$h$ & $0.76^{+0.02}_{-0.02} (0.78)$ & $0.74^{+0.02}_{-0.02} (0.74)$& $0.75^{+0.02}_{-0.02} (0.77)$ & $0.73^{+0.02}_{-0.02}(0.73)$\\
$\Omega_b h^2$ & $0.0216^{+0.0005}_{-0.0005}(0.0222)$ & $0.0216^{+0.0005}_{-0.0006}(0.0221)$ & $0.0216^{+0.0005}_{-0.0006}(0.0221)$ & $0.0217^{+0.0005}_{-0.0006}(0.0223)$ \\
$p$ & $53.18^{+30.63}_{-31.26}(75.80)$ & $54.25^{+30.09}_{-31.01}(63.71)$& $53.99^{+30.97}_{-31.19} (50.23)$ & $53.87^{+30.26}_{-30.43} (48.61)$\\
\hline
$\Omega_m$ & $0.30^{+0.06}_{-0.05}(0.28)$ & $0.32^{+0.06}_{-0.05} (0.32)$ & $0.32^{+0.05}_{-0.05} (0.34)$ & $0.33^{+0.05}_{-0.05} (0.34)$ \\
$n$ & $-0.01^{+0.64}_{-0.63}(-0.35)$ & $0.30^{+0.81}_{-0.77} (-0.01)$ &$0.18^{+0.66}_{-0.66} (0.30)$ & $0.44^{+0.76}_{-0.77}(0.31)$\\
$h$ & $0.72$ & $0.72$& $0.72$ & $0.72$\\
$\Omega_b h^2$ & $0.0216^{+0.0005}_{-0.0007}(0.0220)$ & $0.0216^{+0.0006}_{-0.0006}(0.0222)$ & $0.0216^{+0.0006}_{-0.0005}(0.0223)$ & $0.0217^{+0.0005}_{-0.0005}(0.0223)$ \\
$p$ & $56.14^{+30.28}_{-30.46}(87.19)$ & $54.35^{+30.59}_{-30.26}(61.34)$& $55.19^{+30.44}_{-30.54} (73.54)$ & $54.29^{+30.89}_{-31.33} (73.01)$\\
\hline
$\Omega_m$ & $0.24^{+0.03}_{-0.03}(0.20)$ & $0.27^{+0.05}_{-0.05} (0.23)$ & $0.25^{+0.04}_{-0.04} (0.20)$ & $0.29^{+0.06}_{-0.06} (0.24)$ \\
$n$ & $-0.24^{+0.28}_{-0.30}(-0.48)$ & $-0.04^{+0.46}_{-0.47} (-0.31)$ &$-0.23^{+0.31}_{-0.34} (-0.57)$ & $0.08^{+0.59}_{-0.57}(-0.33)$\\
$h$ & $0.76^{+0.02}_{-0.02} (0.77)$ & $0.74^{+0.03}_{-0.02} (0.76)$& $0.75^{+0.02}_{-0.02} (0.76)$ & $0.73^{+0.02}_{-0.02}(0.75)$\\
$\Omega_b h^2$ & $0.0213$ & $0.0213$ & $0.0213$ & $0.0213$ \\
$p$ & $53.09^{+31.72}_{-31.59}(89.47)$ & $55.33^{+30.73}_{-30.56}(61.07)$& $53.62^{+31.94}_{-32.17} (32.14)$ & $55.03^{+30.79}_{-31.23} (84.36)$\\
\hline
$\Omega_m$ & $0.29^{+0.04}_{-0.05}(0.26)$ & $0.31^{+0.05}_{-0.05} (0.26)$ & $0.30^{+0.05}_{-0.04} (0.27)$ & $0.32^{+0.04}_{-0.05} (0.28)$ \\
$n$ & $-0.19^{+0.52}_{-0.51}(-0.57)$ & $0.01^{+0.72}_{-0.57} (-0.54)$ &$-0.02^{+0.56}_{-0.56} (-0.42)$ & $0.20^{+0.63}_{-0.63}(-0.31)$\\
$h$ & $0.72$ & $0.72$& $0.72$ & $0.72$\\
$\Omega_b h^2$ & $0.0213$ & $0.0213$ & $0.0213$ & $0.0213$ \\
$p$ & $56.95^{+29.88}_{-30.45}(76.57)$ & $55.75^{+29.97}_{-30.44}(78.65)$& $55.84^{+30.50}_{-30.70} (91.07)$ & $55.37^{+30.50}_{-30.49} (96.67)$\\
\hline
$\Omega_m$ & $0.31^{+0.07}_{-0.07}(0.26)$ & $0.37^{+0.07}_{-0.07} (0.36)$ & $0.34^{+0.07}_{-0.07} (0.32)$ & $0.38^{+0.06}_{-0.05} (0.41)$ \\
$n$ & $0.48^{+0.80}_{-0.75}(0.00)$ & $1.28^{+1.16}_{-1.10} (1.14)$ &$0.73^{+0.87}_{-0.86} (0.42)$ & $1.32^{+0.97}_{-0.96}(1.57)$\\
$h$ & $0.76^{+0.02}_{-0.03} (0.78)$ & $0.74^{+0.02}_{-0.02} (0.75)$& $0.74^{+0.03}_{-0.03} (0.76)$ & $0.73^{+0.02}_{-0.02}(0.73)$\\
$(\Omega_b h^2)^*$ & $0.0238^{+0.0010}_{-0.0009}(0.0248)$ & $0.0240^{+0.0008}_{-0.0008}(0.0249)$ & $0.0240^{+0.0008}_{-0.0007}(0.0249)$ & $0.0241^{+0.0007}_{-0.0007}(0.0248)$ \\
$p$ & $52.61^{+31.47}_{-31.78}(34.64)$ & $51.43^{+32.48}_{-32.23}(76.04)$& $51.34^{+32.46}_{-32.52} (37.58)$ & $51.44^{+32.02}_{-32.32} (22.30)$\\
\end{tabular}
\end{ruledtabular}
\end{table*}

\begin{table*}
\caption{\label{Tab5} Mean of the marginal posterior PDF for the parameters of model 4, together with the $68\%$ credible interval, for the SNIa+CMB and SNIa+CMB+BAO data sets. The corresponding values of posterior mode are presented in brackets.}
\begin{ruledtabular}
\begin{tabular}{lcccc}
& SNLS+CMB & SNLS+CMB+BAO & Union+CMB & Union+CMB+BAO \\ \hline
$\Omega_m$ & $0.24^{+0.04}_{-0.03}(0.20)$ & $0.28^{+0.06}_{-0.06}(0.25)$ & $0.26^{+0.05}_{-0.05}(0.23)$ & $0.30^{+0.07}_{-0.06}(0.26)$ \\
$n$ & $-0.32^{+0.29}_{-0.34}(-0.58)$ & $-0.06^{+0.52}_{-0.54}(-0.37)$ &$-0.26^{+0.40}_{-0.40}(-0.56)$ &$0.07^{+0.69}_{-0.65}(-0.38)$\\
$h$ & $0.76^{+0.02}_{-0.03}(0.77)$ & $0.74^{+0.02}_{-0.02}(0.75)$& $0.74^{+0.02}_{-0.02}(0.75)$ &$0.73^{+0.02}_{-0.02}(0.74)$\\
$\Omega_b h^2$ & $0.0216^{+0.0005}_{-0.0006}(0.0220)$ & $0.0216^{+0.0005}_{-0.0006}(0.0221)$ & $0.0216^{+0.0005}_{-0.0006}(0.0222)$ &  $0.0216^{+0.0005}_{-0.0006}(0.0220)$ \\
$m$ & $3.00^{+1.34}_{-1.31}(3.05)$ & $3.20^{+1.23}_{-1.22}(3.18)$& $3.13^{+1.29}_{-1.28}(3.79)$ & $3.22^{+1.22}_{-1.24}(3.61)$\\
\hline
$\Omega_m$ & $0.29^{+0.06}_{-0.05}(0.25)$ & $0.31^{+0.06}_{-0.06}(0.29)$ & $0.30^{+0.06}_{-0.05}(0.30)$ & $0.32^{+0.06}_{-0.06}(0.31)$ \\
$n$ & $-0.24^{+0.58}_{-0.57}(-0.64)$ & $0.04^{+0.79}_{-0.73}(-0.33)$ &$-0.10^{+0.65}_{-0.62}(-0.24)$ &$0.17^{+0.78}_{-0.74}(-0.15)$\\
$h$ & $0.72$ & $0.72$& $0.72$ &$0.72$\\
$\Omega_b h^2$ & $0.0215^{+0.0006}_{-0.0006}(0.0219)$ & $0.0216^{+0.0005}_{-0.0007}(0.0222)$ & $0.0216^{+0.0005}_{-0.0007}(0.0222)$ &  $0.0217^{+0.0005}_{-0.0006}(0.0222)$ \\
$m$ & $3.23^{+1.18}_{-1.16}(4.21)$ & $3.25^{+1.23}_{-1.19}(4.79)$& $3.25^{+1.20}_{-1.16}(4.83)$ & $3.22^{+1.25}_{-1.19}(3.55)$\\
\hline
$\Omega_m$ & $0.24^{+0.03}_{-0.03}(0.20)$ & $0.26^{+0.05}_{-0.04}(0.22)$ & $0.25^{+0.04}_{-0.04}(0.21)$ & $0.28^{+0.07}_{-0.05}(0.22)$ \\
$n$ & $-0.38^{+0.26}_{-0.30}(-0.55)$ & $-0.23^{+0.39}_{-0.43}(-0.50)$ &$-0.36^{+0.33}_{-0.35}(-0.64)$ &$-0.11^{+0.54}_{-0.55}(-0.66)$\\
$h$ & $0.76^{+0.02}_{-0.03}(0.76)$ & $0.74^{+0.02}_{-0.02}(0.75)$& $0.74^{+0.02}_{-0.02}(0.75)$ &$0.73^{+0.02}_{-0.02}(0.74)$\\
$\Omega_b h^2$ & $0.0213$ & $0.0213$ & $0.0213$ &  $0.0213$ \\
$m$ & $3.00^{+1.34}_{-1.33}(4.69)$ & $3.29^{+1.19}_{-1.15}(4.50)$& $3.17^{+1.26}_{-1.28}(4.15)$ & $3.30^{+1.18}_{-1.19}(3.93)$\\
\hline
$\Omega_m$ & $0.28^{+0.04}_{-0.05}(0.24)$ & $0.29^{+0.06}_{-0.05}(0.24)$ & $0.29^{+0.05}_{-0.05}(0.25)$ & $0.30^{+0.06}_{-0.05}(0.26)$ \\
$n$ & $-0.39^{+0.44}_{-0.47}(-0.76)$ & $-0.16^{+0.62}_{-0.58}(-0.73)$ &$-0.27^{+0.52}_{-0.50}(-0.64)$ &$-0.05^{+0.62}_{-0.60}(-0.52)$\\
$h$ & $0.72$ & $0.72$& $0.72$ &$0.72$\\
$\Omega_b h^2$ & $0.0213$ & $0.0213$ & $0.0213$ &  $0.0213$ \\
$m$ & $3.26^{+1.20}_{-1.17}(3.23)$ & $3.33^{+1.16}_{-1.14}(3.66)$& $3.30^{+1.16}_{-1.14}(3.87)$ & $3.30^{+1.19}_{-1.16}(4.33)$\\
\hline
$\Omega_m$ & $0.31^{+0.08}_{-0.07}(0.25)$ & $0.38^{+0.07}_{-0.08}(0.35)$ & $0.34^{+0.08}_{-0.07}(0.34)$ & $0.39^{+0.06}_{-0.06}(0.40)$ \\
$n$ & $0.37^{+0.83}_{-0.79}(-0.22)$ & $1.19^{+1.21}_{-1.16}(0.70)$ &$0.60^{+0.86}_{-0.85}(0.47)$ &$1.25^{+1.01}_{-0.99}(1.43)$\\
$h$ & $0.76^{+0.02}_{-0.03}(0.77)$ & $0.74^{+0.02}_{-0.03}(0.75)$& $0.74^{+0.02}_{-0.02}(0.75)$ &$0.73^{+0.02}_{-0.02}(0.74)$\\
$(\Omega_b h^2)^*$ & $0.0239^{+0.0009}_{-0.0010}(0.0248)$ & $0.0240^{+0.0008}_{-0.0008}(0.0248)$ & $0.0241^{+0.0007}_{-0.0008}(0.0249)$ &  $0.0241^{+0.0007}_{-0.0007}(0.0249)$ \\
$m$ & $2.72^{+1.54}_{-1.52}(3.32)$ & $2.82^{+1.49}_{-1.52}(4.10)$& $2.75^{+1.54}_{-1.54}(3.55)$ & $2.79^{+1.51}_{-1.56}(3.23)$\\
\end{tabular}
\end{ruledtabular}
\end{table*}

Observe that there is weak evidence to favour the $\Lambda$CDM model over model 4 for the all considered cases, besides case 5 and for the SNIa+CMB+BAO data sets, where the comparison is inconclusive. The differences in the values of the evidence between model 4 and model 2 or model 3 are small, which 
leads to the conclusion that there is insufficient information to favour the former. Model 4 is better than model 1 in light of the data used in this analysis, with moderate evidence against the latter. This conclusion is changed in case 5, in which the value of the Bayes factor calculated for those models becomes smaller.

Observe that except for the SNLS+CMB a change in the priors for $h$ and $\Omega_b h^2$ does not change the shape of the posterior PDF for $m$, which is flat for $m>3$. For the SNLS+CMB data, the posterior is wider, it is flat for $m>2$ and becomes narrower when the value of $h$ is fixed. When the range of $\Omega_b h^2$ is extended, the posterior PDF for $m$ becomes wider in all cases. 
\end{itemize}

The above results are encouraging and motivate further study of 
the backreaction models. Especially, it is important to study 
assumptions other than (\ref{ansatz1}) and (\ref{ansatz2}).
As seen, if only assumption (\ref{ansatz2}) was modified (models 2-4)
then not only do we obtained a better fit, but also the value of $\Omega_m$ realistically decreases compared to model 1.

\section{Conclusions}

In this paper we presented a Bayesian analysis of the 
backreaction models.
This work was motivated by the recently proposed 
model of the inhomogeneous alternative to dark energy \cite{LABKC08}.
In this approach the Universe is modeled by the Buchert equations,
which describe the relations between the scale factor, the
average spatial curvature, the average matter distribution, the average expansion and the shear. Larena et al. \cite{LABKC08} showed that their model is consistent with supernova and CMB data. 
Here we included the BAO data and tested this model within
the Bayesian approach (our model 1). 

Our analysis shows that the SNLS and CMB data alone strongly favour the 
$\Lambda$CDM model. With the Union sample and BAO data there is almost strong evidence ($\ln B_{21} > 4$) to favour the $\Lambda$CDM model over 
our model 1.
However, if just the best-fit models are compared, then the $\chi^2$ 
(Union+CMB+BAO) for the $\Lambda$CDM and model 1 are 
$320.96$ and $325.36$ respectively.
If the $\chi^2$ distribution is assumed, then
for 307 degrees of freedom  in model 1 the
probability that this model is true in the light of data is 
22.6\%. In comparison, the $\Lambda$CDM model (308 degrees of freedom) 
gives 29.3\%. Therefore, 
we can see that the best-fit model 1 fits observations almost as well as the $\Lambda$CDM model.
Still, there are other concerns regarding this model.
For example, can the assumptions 
(\ref{ansatz1}) and (\ref{ansatz2}) be justified?
In other words, how does the backreaction and the spatial curvature of the real Universe evolve, and does the relation $\mathcal{Q} \sim
\av{\mathcal{R}}$ for the real Universe hold?
The most concerning is the value of $\Omega_m$, which is quite large, $0.43^{+0.03}_{-0.04}$.
However, as it was shown in Sec. \ref{curidx}, after the assumption 
(\ref{ansatz2}) was modified, we were able to obtain a better fit and, 
in addition, the value of $\Omega_m$ decreased to 
$0.31^{+0.06}_{-0.07}$ and $0.30^{+0.07}_{-0.06}$ (or even lower if the BAO data is excluded -- see Tables \ref{Tab4} and \ref{Tab5})
for models 4 and 5 respectively. 
This shows that a lot still needs to be done in the context
of the backreaction models, especially in the study of the relation
between the average spatial curvature and the backreaction.

We also investigated the sensitivity of the results to the prior assumptions on the $h$, $\Omega_b h^2$ and $n$ parameters. The most prominent changes are in the case in which the Hubble parameter is fixed and in the case in which the range of the parameter $\Omega_b h^2$ is extended. In the latter case, the evidence against model 1 with respect to the $\Lambda$CDM model is weak. 

Currently the $\Lambda$CDM model is preferred by the observational data, but it
is possible that, after the revision of 
assumptions (\ref{ansatz1}) and (\ref{ansatz2}) 
we could obtain a more satisfactory results (see Table~\ref{Tab3}
and Fig.~\ref{fig6}). We
should also remember that in these models of dark energy,
the dark-energy-term appears as a consequence 
of  inhomogeneities that are present in the Universe. Therefore,
within this class of models, the ``decaying lambda term'' takes on reveals a new and natural interpretation.

\section*{Acknowledgments}

This research was supported by the Marie Curie Host Fellowships for the Transfer of Knowledge project COCOS (Contract No. MTKD-CT-2004-517186) (AK, MS)
and partly by the Peter and Patricia Gruber Foundation
and the International Astronomical Union (KB). MS is very grateful to prof. Mauro Carfora for discussion and warm atmosphere during the visit in Pavia.
We also thank Andrea Prinsloo for helpful comments.


\section*{References}


\begin{thebibliography}{99}

\bibitem{branes-review}
K.~Koyama, {Gen. Rel. Grav.} {\bf 40}, 421 (2008).

\bibitem{f(R)-review}
S.~Capozziello and M.~Francaviglia, {Gen. Rel. Grav.} {\bf 40}, 357 (2008).

\bibitem{icm-review}
M.N.~C\'el\'erier, {New Adv. Phys.} {\bf 1}, 29 (2007).


\bibitem{B00} 
T.~Buchert, {Gen. Rel. Grav.} {\bf 32}, 105 (2000).

\bibitem{R06} 
S.~R\"as\"anen, {J. Cosmol. Astropart. Phys.} {\bf 11}, 003 (2006).

\bibitem{B08} 
T.~Buchert, {Gen. Rel. Grav.} {\bf 40}, 467 (2008).

\bibitem{BC08}
T.~Buchert and M.~Carfora, {Class. Quant. Grav.} {\bf 25}, 195001 (2008).

\bibitem{LABKC08} 
J.~Larena, J.-M.~Alimi, T.~Buchert, M.~Kunz, and P.-S.~Corasaniti,
{Phys. Rev.} {\bf D79}, 083011 (2009).

\bibitem{PS08}
A.~Paranjape and T.P.~Singh, {Gen. Rel. Grav.} {\bf 40}, 139 (2008).


\bibitem{ES87} 
G. F. R. Ellis and W. Stoeger,
{\it Class. Quant. Grav}. {\bf 4}, 1697 (1987).


\bibitem{Mukherjee:2005wg}
P.~Mukherjee, D.~Parkinson, and A.R.~Liddle, {Astrophys. J.} \textbf{L51}, 638 (2006); code available from http://www.cosmonest.org/

\bibitem{Skilling}
J.~Skilling, {\it Bayesian Inference and Maximum Entropy Methods in Science and Engineering} {AIP Conf. Proc. vol 735 p~395} (2004).; http://www.inference.phy.cam.ac.uk/bayesys/


\bibitem{Lewis:2002ah}
A.~Lewis and S.~Bridle, {Phys. Rev.} \textbf{D66}, 103511 (2002);
code available from http://cosmologist.info/cosmomc/


\bibitem{J61}
H.~Jeffreys, {\it Theory of Probability} (Oxford University Press, Oxford, 1961).


\bibitem{A06}
P.~Astier {\em et al}., {Astron. Astrophys}. {\bf 447}, 31 (2008).

\bibitem{K08} 
M.~Kowalski {\em et al}., arXiv:0804.4142 (2008).

\bibitem{DL02}
M.~Doran and M.~Lilley, {Mon. Not. R. Astron. Soc.} \textbf{330}, 965 (2002).

\bibitem{WMAP3} 
G.~Hinshaw {\em et al}., {Astrophys. J. Suppl. Ser.} \textbf{170}, 288 (2007).

\bibitem{Hu96}
W.~Hu and N.~Sugiyama, {Astrophys. J. } \textbf{471}, 542 (1996).

\bibitem{Freedman01}
W.L.~Freedman {\em et al}., {Astrophys. J.} \textbf{553}, 47 (2001).

\bibitem{Pettini08}
M~Pettini {\em et al}., arXiv:0805.0594 (2008). 

\bibitem{RT08}
R.~Trotta,  {Contemporary Physics} \textbf{49}, 71 (2008). 

 \end{thebibliography}
\end{document}